\documentclass[10pt,preprint]{aastex}

\begin{document}

\title{NITROGEN AND OXYGEN ABUNDANCE VARIATIONS IN THE OUTER EJECTA OF
ETA CARINAE: EVIDENCE FOR RECENT CHEMICAL ENRICHMENT}

\author{Nathan Smith\altaffilmark{1,2} and Jon A.\ Morse\altaffilmark{3}}

\affil{Center for Astrophysics and Space Astronomy, University of
Colorado, 389 UCB, Boulder, CO 80309}


\altaffiltext{1}{Hubble Fellow; nathans@casa.colorado.edu}
\altaffiltext{2}{Visiting Astronomer, Cerro Tololo Inter-American
Observatory, National Optical Astronomy Observatory, operated by the
Association of Universities for Research in Astronomy, Inc., under
cooperative agreement with the National Science Foundation.}
\altaffiltext{3}{Current Address: Department of Physics \& Astronomy,
Arizona State University, Box 871504, Tempe, AZ 85287-1504}

\begin{abstract}

We present optical spectra of the ionized `Outer Ejecta' of $\eta$
Carinae that reveal differences in chemical composition at various
positions.  In particular, young condensations just outside the dusty
Homunculus Nebula show strong nitrogen lines and little or no oxygen
--- but farther away, nitrogen lines weaken and oxygen lines become
stronger.  The observed variations in the apparent N/O ratio may
signify either that the various blobs were ejected with different
abundances, or more likely, that the more distant condensations are
interacting with normal-composition material.  The second hypothesis
is supported by various other clues involving kinematics and X-ray
emission, and would suggest that $\eta$ Car is enveloped in a
``cocoon'' deposited by previous stellar-wind mass loss.  In
particular, all emission features where we detect strong oxygen lines
are coincident with or outside the soft X-ray shell.  In either case,
the observed abundance variations suggest that $\eta$ Car's ejection
of nitrogen-rich material is a {\it recent} phenomenon --- taking
place in just the last few thousand years.  Thus, $\eta$ Carinae may
be at a critical stage of evolution when ashes of the CNO cycle have
just appeared at its surface.  Finally, these spectra reveal some
extremely fast nitrogen-rich material, with Doppler velocities up to
3200 km s$^{-1}$, and actual space velocities that may be much higher.
This is the fastest material yet seen in $\eta$ Car's nebula, but with
unknown projection angles its age is uncertain.

\end{abstract}


\section{INTRODUCTION}

In the hot interiors of massive stars, the equilibrium CNO cycle
converts most of the carbon and oxygen into nitrogen.  Through
turbulent and rotational mixing (Maeder 1982), or perhaps by stripping
of the outer hydrogen-rich envelope by a stellar-wind, these
nitrogen-rich ashes from the core may eventually be observed in a
massive star's atmosphere or its ejecta.  One spectacular example of
N-rich ejecta in a massive star's circumstellar nebula, in addition to
the rings around SN 1987a (Sonneborn et al.\ 1997; Meaburn et al.\
1995), is the `Outer Ejecta' around the evolved massive star $\eta$
Carinae.

Outside its dusty bipolar reflection nebula known as the `Homunculus',
which was ejected in the mid-19th century, $\eta$ Carinae is
surrounded by a complex aggregate of ionized gas condensations known
collectively as the `Outer ejecta', the `Outer Shell', or the `Outer
Condensations' (e.g., Walborn 1976; Thackeray 1950; Meaburn et al.\
1996).  Since these ejecta are ionized, as opposed to mostly neutral
and dusty like the Homunculus, they provide a practical way to
estimate the chemical abundances of some material recently ejected by
$\eta$ Carinae.  Some of the more prominent nebular condensations in
the outer ejecta have names that are identified in the {\it HST}/WFPC2
image in Figure 1 (see Morse 1999; Morse et al.\ 1998), following the
nomenclature of Walborn (1976).  Proper motions suggest that some of
these condensations may predate $\eta$ Car's famous 19th century
eruption by a few hundred years or more (Walborn et al. 1978, Walborn
\& Blanco 1988), while some ejecta closer to the Homunculus originated
within a few decades of that eruption (Morse et al.\ 2001).  Bright
ionized condensations in the outer ejecta coincide with a soft X-ray
emitting shell (Seward et al.\ 2001; Chlebowski et al.\ 1984),
suggesting that they are excited predominantly by shocks, rather than
photoionized.  Thus, young fast ejecta may be plowing into older
material.  Davidson et al.\ (1982, 1986) examined the UV and optical
spectrum of the brightest condensation and the brightest X-ray source
in the outer ejecta (the `S Condensation') and found it to be
extremely nitrogen rich, being almost completely devoid of any oxygen
or carbon lines.  Thus, ashes of the CNO cycle have been exposed at
the star's surface, confirming that $\eta$ Car is indeed an evolved
massive star. Dufour (1989) and Dufour et al.\ (1997) have also
discussed the physical properties and abundances in the S Condensation
and some similar nearby ejecta.  However, no suitable spectra for
examining abundances of other condensations in the Outer Ejecta have
been published; consequently, it is usually assumed that all of the
Outer Ejecta are nitrogen rich.

Here we present long-slit optical spectra for several positions in
$\eta$ Car's Outer Ejecta.  We find that the chemical abundances are
{\it not} uniform, and we discuss interpretations and evolutionary
implications of this conclusion.  We present our spectroscopic
observations in \S 2 and discuss the data in \S 3, including a
deductive chemical abundance analysis.  In \S 4 we discuss
implications of the non-uniform abundances in the young ejecta around
$\eta$ Car, as well as some interesting details of the observed
kinematics.

\section{OBSERVATIONS}

Low resolution ($R \sim$ 700-1600; 2-pixel) spectra from 3600 to 9700
\AA \ were obtained on 2002 March 1 and 2 using the RC Spectrograph on
the CTIO 1.5-m telescope.  The 7$\arcmin$ long and 1$\farcs$5 wide
slit aperture was oriented at position angle 310$\arcdeg$, and pointed
at two different positions in the nebula around $\eta$ Car as shown in
Figure 1.  Spectra at each pointing were obtained on two separate
nights in two different wavelength ranges (blue: 2600 - 7100 \AA, and
red: 6250 - 9700 \AA), with total exposure times of 1200 s each.  Sky
conditions were photometric; flux-calibration and telluric absorption
correction (at far-red wavelengths) were accomplished using similar
observations of the standard stars LTT-3218 and LTT-2415, with airmass
correction performed using the CTIO extinction coefficients in {\tt
IRAF}.  We estimate that the absolute photometric accuracy of our
spectroscopic data is roughly $\pm$5 to 10\% over most of the observed
wavelength range, increasing to between 10 and 20\% toward the blue
end of the spectrum. Note, however, that our analysis below is based
exclusively on {\it relative} line intensities.

In the resulting two-dimensional (2-D) spectra, various bright
condensations corresponding to ejecta seen in Figure 1 -- some with
very high Doppler shifts -- could be seen superimposed on narrow
emission lines from the Carina Nebula H~{\sc ii} region (and some
faint narrow sky lines) with constant velocity along the entire length
of the slit.  We subtracted this relatively smooth background emission
by carefully fitting the spectrum on each side of each condensation
and interpolating with a second-order polynomial, excluding the bright
ejecta from the fit.  A careful subtraction like this is critical for
interpreting the extracted spectra, and our subtraction method is
reliable excluding unresolved variations in the flux from the H~{\sc
ii} region on size scales of a few
arcseconds.\footnotemark\footnotetext{However, [S~{\sc iii}]
$\lambda$9069 showed some subtraction artifacts (especially in the
fainter W2 and W Edge spectra in Figure 2), which may be partially
caused by telluric absorption.}  The spatial scale of a detector pixel
was 1$\farcs$3.

From these background-subtracted 2-D spectra we then made several
one-dimensional (1-D) tracings of individual condensations; the smaller
boxes in Figure 1 represent the extracted segments of the slit
aperture for each feature.  Names for various features are adopted
from Walborn (1976) and Meaburn et al.\ (1993).  The blue and red
wavelength ranges of these extracted 1-D spectra were then merged to
form a single 3600-9700 \AA \ spectrum for each feature with a common
dispersion of 2 \AA \ pixel$^{-1}$.  The average of the blue and red
segments was taken in the wavelength range near H$\alpha$ where the
two overlapped.  The fluxes of emission lines and faint continuum
agreed to within the noise of the data in this overlapping region,
suggesting that the flux calibration was reliable from night to night.
Figure 2 shows the resulting 1-D optical spectra for several individual
condensations through apertures identified with boxes in Figure 1, as
well as the average H~{\sc ii} region spectrum sampled a few
arcminutes away on either side of $\eta$ Car.  The size of the
extraction region is also indicated in Figure 2 for each tracing.  The
spectra probed emission from [O~{\sc ii}] $\lambda$3727 in the blue
out to [S~{\sc iii}] $\lambda$9532 in the red.

Observed intensities of many relevant lines are listed in Table 1,
relative to H$\beta$=100.  Uncertainties in these line intensities
vary depending on the strength of the line and the measurement method.
Isolated emission lines were measured by taking the integrated flux of
the line; for these, brighter lines with $I>$10 typically have
measurement errors of a few percent, and weaker lines may have
uncertainties of $\pm$10 to 15\%.  In spectra with the lowest
signal-to-noise, like W2, uncertainties are typically $\pm$10\% for
bright lines, increasing to $\pm$25\% for faint lines.  In all cases,
the uncertainties increase somewhat at the blue edge of the spectrum.
Blended pairs or groups of lines were measured by fitting Gaussian
profiles.  For brighter blended lines like H$\alpha$+[N~{\sc ii}] and
[S~{\sc ii}], the measurement uncertainty is typically 5 to 10\%.
Obviously, errors will be on the high end for faint lines adjacent to
bright lines, and errors will be on the low end for the brightest
lines in a pair or group, or lines in a pair with comparable intensity
(like [S~{\sc ii}]).  For faint blended lines, the measurement
uncertainties increase; in Table 1, lines with measurement uncertainty
of more than about $\pm$25 to 30\% are listed in parentheses.

\section{RESULTS AND ANALYSIS}

\subsection{Comments on Individual Spectra}

{\it S Condensation}.  The S Condensation appears to be a position in
the Outer Ejecta where young material from the Great Eruption is
catching-up with older ejecta in the S Ridge (Morse et al.\ 2001).  An
upper limit to [O~{\sc iii}] $\lambda\lambda$4949,5007 is given in the
tables.  A fairly bright line near 5020 \AA \ is seen, but its
centroid velocity is 300 km s$^{-1}$ redder than the Doppler velocity
of other lines in the S Condensation, and it is probably the N~{\sc
ii} $\lambda$5011 blend or perhaps [Fe~{\sc ii}] $\lambda$5018 instead
of [O~{\sc iii}] $\lambda$5007.  Therefore, the [O~{\sc iii}] upper
limits listed in the tables are based upon a measured upper limit for
[O~{\sc iii}] $\lambda$4949 and a limit for [O~{\sc iii}]
$\lambda$5007 that is 3 times stronger.  An upper limit for [O~{\sc
iii}] $\lambda$4363 is not given because the strong line near the
expected wavelength is probably [Fe~{\sc ii}] $\lambda$4358.  The
[O~{\sc i}] lines at 6300 and 6364 \AA \ are uncertain because they
are blended with nearby lines like [S~{\sc iii}] and Si~{\sc ii}, but
the intensities of these lines are probably valid to within about
30\%.  We chose to identify the emission feature at 7325 \AA \ as
[Ca~{\sc ii}] instead of [O~{\sc ii}], due to the strength of [Ca~{\sc
ii}] $\lambda$7291.  The red spectrum of the S Condensation is similar
to spectra observed in Herbig-Haro objects (e.g., HH~47A; Morse et
al.\ 1994), with a wide range of ionization and strong low-excitation
forbidden lines of S, N, Fe, Ca, Ni, etc., produced in
moderate-velocity radiative shock waves.  [Fe~{\sc ii}] $\lambda$8617
is particularly strong, and is characteristic of dense shocks.

{\it S Ridge}.  Material in the S Ridge was probably ejected by the
star at least several decades before the Great Eruption (Walborn et
al.\ 1978; Morse et al.\ 2001). Comments made above regarding [O~{\sc
iii}] and [O~{\sc i}] lines in the S Condensation apply here as well.
In addition, [O~{\sc ii}] $\lambda\lambda$3727,3729 is somewhat
uncertain because it is partially blended with the nearby hydrogen
lines H13 and H14, and again, the line at 7325 \AA \ is probably
[Ca~{\sc ii}] instead of [O~{\sc ii}]. In general, the spectrum of the
S Ridge is similar to that of the S and ES Condensations (Dufour et
al.\ 1997), except that hydrogen lines are relatively stronger here.
It is perhaps interesting to note that for the S Condensation,
hydrogen lines were also relatively stronger than forbidden lines in
the spectrum published by Davidson et al.\ (1986), which was obtained
with a larger aperture than we used for the S Condensation.

{\it NN Jet}.  Like the S Condensation, the NN Jet appears to be
younger than the rest of the Outer Ejecta; it probably originated in
the Great Eruption (Morse et al.\ 2001). The spectrum of the NN Jet is
dominated by reflected light from $\eta$ Carinae itself (notice the
continuum level in the NN Jet spectrum in Figures 2$c$ and 4),
indicating a high dust content in the NN Jet, with a clear
line-of-sight to the star.  Interestingly, there is little or no soft
X-ray emission from the position of the NN Jet in recent {\it Chandra}
images (Seward et al.\ 2001; we return to this mysterious absence of
X-ray emission later).  Some relatively weak and blueshifted intrinsic
emission from ionized gas can be seen in the 2-D spectra (Figure 4),
apparently from the outer bow shock around the ``jet''.  This emission
has strong nitrogen lines and a spectrum qualitatively similar to the
S Ridge, with little or no [O~{\sc ii}] and [O~{\sc iii}] emission.
However, the kinematics of the ionized gas are complicated, precluding
the measurement of accurate line intensities in the low-resolution
spectra presented here (for instance, reflected H$\alpha$ is blended
with [N~{\sc ii}] $\lambda$6583).  Thus, the spectrum of the NN Jet is
not included in the discussion below.  However, Meaburn et al.\ (1996)
have already discussed the kinematics and emission properties of the
NN Jet in considerable detail; their high-resolution spectra show that
the NN Jet is nitrogen rich, with [N~{\sc ii}] $\lambda$6583 roughly 6
times stronger than H$\alpha$ (see also Dufour et al.\ 1997).

{\it W Condensation (W2)}.  The W Condensation identified by Walborn
(1976) was not included in either slit position, but a nearby feature
that we call ``W2'' was included in the northeast slit position (see
Figure 1).  The blob is faint and the observed spectrum is noisy, but
it is useful because it shows the relative strengths of [O~{\sc iii}]
$\lambda\lambda$4949,5007, [N~{\sc ii}] $\lambda\lambda$6548,6583, and
hydrogen lines at this position.  [O~{\sc iii}] $\lambda$4363 and
[S~{\sc iii}] $\lambda$6312 were too faint to measure, so only the
[N~{\sc ii}] electron temperature can be deduced here.  This spectrum
is complicated by multiple velocity components -- the [N~{\sc ii}]
lines are clearly redshifted by $\ga$100 km s$^{-1}$ with respect to
strong hydrogen and oxygen lines (see the inset in Figure 2$d$, for
example, and Figure 4).  These two components may trace different
parts of a shock (see \S 4.2), or perhaps a mixture of ejections.  The
redshifted emission is qualitatively similar to that of the E
Condensation.

{\it E Condensation (E5)}.  One of our two slit positions passed
through a few of the E Condensations identified by Walborn (1976), of
which E5 is the brightest (see Figure 1).  Like the W Condensation,
the E Condensation has several velocity components.  The spectrum here
is dominated by blueshifted gas at a few hundred km s$^{-1}$, but
there is both slower (nearly rest wavelength) and much faster material
as well -- see the discussions about kinematics in \S 4.2 and \S
4.3. As in the W Condensation, O lines arise in the slowest component.
[O~{\sc i}] $\lambda$6300 and [S~{\sc iii}] $\lambda$6312 were
blended, but appeared to have comparable intensities.  The E
Condensation is part of the soft X-ray shell seen in Chandra images
(see Figure 3).

{\it W Edge}.  The southwest slit passed through a thin filament at
the western edge of $\eta$ Car's outer ejecta, which has not been
discussed before.  This feature was visible in some emission lines in
the raw 2-D spectra, and the background spectrum extrapolated from
either side of the aperture marked in Figure 1 was subtracted.  The
resulting spectrum is similar to the background H~{\sc ii} region,
except for stronger [O~{\sc ii}] $\lambda\lambda$3727,3729, an
indication of a slow to moderate velocity shock in an ionized medium
(e.g., see the discussion of HH~47D by Morse et al.\ 1994).  The
radial velocity of the W Edge is similar to the H~{\sc ii} region to
within $\pm$50 km s$^{-1}$.  We identify the 7325 \AA \ feature as the
[O~{\sc ii}] doublet instead of [Ca~{\sc ii}], because [Ca~{\sc ii}]
$\lambda$7291 is not detected.

{\it Carina Nebula}.  The average spectrum of the background Carina
Nebula was measured in 30$\arcsec$-wide extractions of bright areas on
either side of $\eta$ Car's Outer Ejecta (the full slit aperture was
roughly 7$\arcmin$ long; much longer than shown in Figure 1).  Sky
subtraction for these areas was performed by subtracting the spectrum
in relatively dark areas of the nebula, like the dark patches of
extinction in the Keyhole Nebula (Smith 2002; Brooks et al.\ 2000).
Relative line intensities from the background nebula varied along the
slit, but the spectrum in Figure 2$g$ is a suitable average
representing the H~{\sc ii} region in $\eta$ Car's vicinity.

\subsection{Reddening}

We used the observed Balmer decrement in the adjacent Carina Nebula to
determine the line-of-sight reddening toward $\eta$ Carinae.  Using
the extinction law of Cardelli, Clayton, \& Mathis (1989) and Case B
values for hydrogen line intensities computed by Hummer \& Storey
(1987), the H$\beta$ and H$\gamma$ lines in the Carina Nebula
suggested $E(B-V)\approx$0.2$\pm$0.5.  This is lower than the average
value of $E(B-V)$=0.47 for O stars in the Tr16 cluster, but it is
reasonable given the observed range of 0.25 to 0.64 for those same
stars (Walborn 1995).  This value is low compared to values adopted in
previous studies of the S Condensation (0.4 to 0.6; Davidson et al.\
1986; Dufour et al.\ 1997); for this reason, we do not include the
[S~{\sc iii}] electron temperature in the analysis below, since it
covers a fairly wide wavelength range and is more susceptible to
errors from an uncertain reddening correction than the other $T_e$
diagnostics.  (The [S~{\sc iii}] lines yielded temperatures of 8,000
to 12,000 K.)  Note that $R = A_V / E(B-V)$ is not the usual value of
$\sim$3.1; dust clouds around the Keyhole Nebula near $\eta$ Car
suggest values closer to $R \approx 4.8$ (Smith 2002; Smith 1987).
Although there may be variable extinction across $\eta$ Car's
environment, it is hard to characterize accurately with available
data, so we used the same value of $E(B-V)$ = 0.2 to correct line
intensities for all the condensations; dereddened intensities are
listed in Table 2.  Although the H$\alpha$/H$\beta$ ratio varies
considerably (as expected when collisions in shock waves enhance the
H$\alpha$/H$\beta$ ratio; e.g., Chevalier \& Raymond 1978), the
resulting dereddened H$\gamma$/H$\beta$ ratios are within 5-10\% of
the Case B value, indicating that variable extinction is probably not
severe.  In any case, an incorrect $E(B-V)$ would not undermine the
main result of this study: the systematic change in O and N line
strengths at various positions.

\subsection{Deductive Nebular Analysis}

Given the complex kinematics and the non-equilibrium shock physics
that may dominate emission from $\eta$ Car's Outer Ejecta, it is
dangerous to place too much weight on a standard deductive nebular
analysis.  Nevertheless, it is a place to start; the results can guide
inputs for future shock models and may be instructive and useful if we
keep in mind that they apply to the average properties of the observed
condensations.  Table 2 lists electron densities for various
condensations derived from the [S~{\sc ii}]
$\lambda$6717/$\lambda$6731 ratio, and characteristic electron
temperatures derived from the usual [N~{\sc ii}] and [O~{\sc iii}]
line ratios.  These were calculated using the {\tt TEMDEN} task in
{\tt IRAF}'s {\tt NEBULAR} package (see Shaw \& Dufour 1994).  For the
S Condensation, which has already been studied extensively, our value
of $T_{\rm e}$ is somewhat lower than previous estimates, but $n_{\rm
e}$ is reasonably consistent with Davidson et al.\ (1986) and Dufour
et al.\ (1997), especially considering the variations in these
quantities on small spatial scales within the S Condensation itself
(Dufour et al.\ 1997).

A clear trend is seen in Figure 2; namely, from top to bottom,
nitrogen lines weaken systematically as oxygen lines get stronger, and
the characteristics of the spectra change from a rich shock spectrum
to a photoionized H~{\sc ii} region.  Since we do not have
corresponding UV spectra for all observed features, and consequently,
we lack information on critical ionization stages like N$^{++}$ and
higher, it is not immediately clear whether this trend is due to
ionization or an abundance gradient.  Using the temperatures and
densities in Table 2, some crude estimates of the N and O abundances
are possible, however.  Ionic abundances for a few stages of N and O
are given in Table 2, computed using the {\tt NEBULAR} package in {\tt
IRAF}, as well as the relative nitrogen to oxygen ratio given by
[N$^0$+N$^+$]/[O$^0$+O$^+$+O$^{++}$] (this is not the total N/O ratio,
just a few observed ionization stages).  We see that N/O changes from
$\sim$20 in the S Condensation to much lower values $\la$1 in the
condensations farther from the star.

Still, how can we be certain that we are not being misled by changes
in ionization rather than chemical abundances?  Indeed, there appears
to be a pronounced ionization gradient in the outer ejecta: The ratio
[S~{\sc iii}] $\lambda$9069+$\lambda$9532 $\div$ [S~{\sc ii}]
$\lambda$6717+$\lambda$6731 is a rough gauge of the ionization
conditions in the gas which is emitting [O~{\sc iii}], [O~{\sc ii}],
and [N~{\sc ii}], provided that the electron densities do not approach
the [S~{\sc ii}] critical density.  (In Table 2, $n_e$ values for
various regions are all $\la$10$^4$ cm$^{-3}$, so it is unlikely that
[S~{\sc ii}] or [O~{\sc ii}] $\lambda\lambda$3726,3729 are severely
quenched.) In the S Condensation we measure [S~{\sc iii}]/[S~{\sc ii}]
= 0.67, and the ratio increases outward with values of 3.5 (S Ridge),
7.6 (W Cond.\ 2), 7.3 (E Cond.\ 5), and 12 (W Edge). Thus, there is an
ionization gradient toward higher average ionization with distance
from the star, regardless of whether the ionization is from shocks or
photoionization.  From spectroscopy of shocks in HH objects (e.g.,
HH~47D; Morse et al.\ 1994), the relative strength of [O~{\sc i}] and
[O~{\sc iii}] lines are anticorrelated as a function of ionization,
while [O~{\sc ii}] remains fairly constant.  Thus, the oxygen
depletion in the S Condensation may be best noted by the lack of
[O~{\sc i}] $\lambda$6300 emission and very weak [O~{\sc ii}]
$\lambda\lambda$3726,3729, in addition to the absence of [O~{\sc iii}]
$\lambda$5007.

Previous study of the S Condensation, including UV wavelengths that
are not included here, showed strong lines of N$^{+2}$, N$^{+3}$, and
N$^{+4}$, but no strong oxygen lines of higher excitation (Davidson et
al.\ 1982, 1986; Dufour et al.\ 1997), and firmly established that the
S Condensation is indeed N-rich and severely depleted of O and C.  By
comparing our results for the S Condensation, we can check the
validity of our abundance analysis.  In general, our N and O
abundances are consistent with these earlier studies of the S
Condensation.  For example, for the N abundance on a logarithmic scale
with H=12, we find [(N$^0$+N$^+$)/H]=8.6, whereas Dufour et al.\
(1997) find [(N$^+$+N$^{2+}$+N$^{3+}$+N$^{4+}$)/H]=8.78.  Similarly,
we find [(O$^0$+O$^+$+O$^{2+}$)/H]=7.3, and Dufour et al.\ find
[(O$^+$+O$^{2+}$)/H]=7.06.  The agreement is good, considering the
different ionization levels included and the difference in aperture
sizes (Dufour et al.\ 1997 used {\it HST} data).  In any case, despite
potential uncertainties from calibration, reddening correction, or
other systematic effects, it is clear from Figure 2 that the W
Condensation, the E Condensation, and the W Edge are not severely
depleted of oxygen, and it is likely that the observed variation in
the N/O ratio from one condensation to another is a real effect of
chemical abundances.

\section{DISCUSSION}

\subsection{Possible Interpretations}

Our observations suggest inhomogeneous abundances in the Outer Ejecta
of $\eta$ Carinae.  Specifically, there is a strong change in the
apparent N/O ratio: while ejecta immediately outside the Homunculus
are nitrogen rich and severely oxygen depleted, more distant material
appears to have normal abundances.  If it is indeed true, how shall we
interpret this apparent chemical abundance gradient?  Two different
scenarios seem like obvious possibilities:

1.  Dense condensations ejected by $\eta$ Carinae over the past few
    thousand years may have become progressively more enriched with
    nitrogen and depleted of oxygen as ashes from the CNO cycle
    appeared at the surface of the star, or as mass loss ate deeper
    into the processed material inside the star.

2.  All of the dense condensations that make up the Outer Ejecta may
    indeed be nitrogen rich (perhaps in varying degrees), but the
    older material farther from the star may be interacting with the
    normal composition material deposited by previous stellar-wind
    mass loss.  This would imply that $\eta$ Car's Outer Ejecta are
    expanding into a cocoon or halo created by the loss of the star's
    original hydrogen-rich stellar envelope, as depicted in Figure 5.

Observed kinematics may help distinguish between these possibilities.
For example, proper motion measurements may be able to determine if
the ejecta are expanding freely or are being decelerated, and ejecta
with different chemical abundances may show discrepant Doppler
velocities.  Higher-quality data are needed for conclusive answers
(high-dispersion spectroscopy of oxygen lines might be particularly
useful), but some more immediate clues are discussed below.
Proper-motion measurements in the S Ridge (Walborn \& Blanco 1988;
Morse et al.\ 2001) show a very large scatter in the projected
velocities (about 5 times larger than the Homunculus), suggesting
interactions between slower condensations and faster ejecta or a wind.

\subsection{Clues from Kinematics}

Of the two possibilities listed above, the second appears to be
supported more closely by the observations. A prominent soft X-ray
shell surrounds $\eta$ Car, and the shock heating of this shell arises
as fast ejecta overtake slower material.  Figure 3 shows contours of
the soft X-ray emission observed by {\it Chandra} ACIS-I (see Seward
et al.\ 2001), superposed on an optical [N~{\sc ii}] image of the
Outer Ejecta taken in June 1999 with {\it HST}/WFPC2.  Condensations
that show relatively bright oxygen lines in their spectra in Figure 2
(E5, W2, W Edge) coincide with or lie outside the soft X-ray shell,
whereas oxygen-poor ejecta (S Ridge, NN Jet) are inside it.  The S
Condensation, S Ridge, and NN Jet may comprise an `inner shell'
(Meaburn et al.\ 1996) that may be distinct from more distant ejecta
(note, however, that the S Condensation and NN Jet appear to have been
ejected in the Great Eruption, while the S Ridge originated somewhat
earlier; Morse et al.\ 2001).  Thus, if one wishes to estimate
chemical abundances of $\eta$ Car pertaining to its current
evolutionary state, one must study the ionized ejecta of this `inner
shell' which is not yet contaminated by swept-up material.

Examining long-slit spectra like the detail of [N~{\sc ii}] and
H$\alpha$ lines shown in Figure 4 provides some additional clues.
While the background H~{\sc ii} region emission has been subtracted
carefully, Figure 4 shows some faint H$\alpha$ emission near the
systemic velocity within $\sim$30$\arcsec$ of $\eta$ Car.  This
implies that the Outer Ejecta are embedded in a cocoon or halo of
H-rich material.  Even with a careful subtraction of the spectrum on
either side of a given condensation like E5, there is still some
enhanced low-velocity H$\alpha$ emission at the position of the
condensation that contaminates the spectrum in Figure 2.  This may
indicate some low-velocity H-rich gas associated with the E5
condensation itself; the oxygen lines in the E5 condensation also
appear to have lower velocities than the nitrogen lines.  Obviously
our analysis is hampered by low spectral resolution, so it is
difficult to confirm any direct relation between kinematics and
chemical abundances.  However, Meaburn et al. (1996) presented echelle
spectra of H$\alpha$ and the [N~{\sc ii}] lines for several
condensations and showed a clear trend: the [N~{\sc ii}]/H$\alpha$
ratio was stronger in fast ejecta just outside the Homunculus, and
dropped by a factor of $\sim$2 in the slower ejecta farther from the
star.  Thus, even at high dispersion where line profiles are resolved,
line intensities may be qualitatively consistent with an abundance
gradient.

\subsection{Extremely Fast Ejecta}

Figure 4$a$ also reveals some extremely fast nitrogen-rich ejecta not
previously reported.  At one slit position northeast of the star, we
see blueshifted emission up to about $-$3200 km s$^{-1}$ and
redshifted emission as fast as $+$2000 km s$^{-1}$.  The extremely
fast blueshifted material is faint, and was not seen in previous
investigations (Meaburn et al.\ 1987, 1993, 1996; Dufour 1989; Weis et
al.\ 2001).  Dufour (1989) detected a redshifted emission feature at
$+$2150 km s$^{-1}$, but no blueshifted features faster than $-$1000
km s$^{-1}$ were reported.

The fastest blueshifted emission in Figure 4$a$ is particularly
interesting.  First, this material is nitrogen-rich and the Doppler
shifts up to $-$3200 km s$^{-1}$ are probably correct, since the two
nitrogen lines and perhaps even H$\alpha$ can be seen offset from one
another by the expected amount (these features are identified as
``EFE'' for ``extremely fast ejecta'' in Figure 4).  Coincidentally,
this fast blueshifted emission extends spatially away from the
Homunculus as far as the position of the E Condensation 5 (roughly
$-$25$\arcsec$ along the slit in Figure 4), where it seems to end
abruptly.  {\it Is this fast material colliding with the E
Condensation?}  If so, there are interesting implications for the
geometry.  The E Condensation 5 has a trajectory tilted out of the
plane of the sky by $\alpha \approx 13\arcdeg$, since it has a Doppler
shift of $-$140 km s$^{-1}$ (Meaburn et al.\ 1996) and a tangential
speed from proper motions of $\sim$600 km s$^{-1}$ (Walborn et al.\
1978).  If the same value of $\alpha$ applies to the EFE, the true
space velocity away from $\eta$ Car would be well above 10$^4$ km
s$^{-1}$.  Such astonishingly high speeds would be unusual even for
young supernova remnants (e.g., Fesen et al.\ 1988).  Though $\alpha$
is uncertain, the true space velocity is likely to be well above 3200
km s$^{-1}$, since the material is certainly not coming directly
toward us -- the fastest material is seen $\sim$25$\arcsec$ away from
the star. {\it With speeds well above 3200 km s$^{-1}$, this is the
fastest nebular material yet detected that is associated with $\eta$
Car.}

Could this `EFE' signify a shock from the Great Eruption that is now
causing the X-ray shell around $\eta$ Car?  Obviously, better
spectroscopic data are desirable to constrain the age of this
interesting fast ejecta. Unfortunately, the proper motion of this
emission cannot be measured with existing {\it HST} data because it is
Doppler shifted out of narrow imaging filters centered on H$\alpha$ or
[N~{\sc ii}].  Regardless of the age, it is likely that this fast
material was ejected by $\eta$ Car itself (i.e. the evolved primary in
the putative binary system that was responsible for the Great
Eruption), because it is nitrogen rich, with [N~{\sc ii}]
$\lambda$6548 $\div$ H$\alpha$ more than 6.  This indicates that under
some circumstances, the primary star is indeed capable of ejecting
material at speeds sufficient to account for the hard X-ray emission
that varies with $\eta$ Car's 5.5 year cycle (Pittard \& Corcoran
2002).  Such high speeds are not seen in spectroscopy of the primary
star's stellar wind, even at the poles where the outflow speed is
highest (Smith et al.\ 2003a).

If this fast material is sweeping-out a cavity and causing the soft
X-ray emission shell in Figure 3, it may help to explain why there is
no significant X-ray emission coming from the NN Jet and some other
Outer Ejecta to the north and east of the Homunculus -- i.e. even
though the NN Jet is moving very fast, it is plowing through material
that is already moving outward at high speed, so the relative shock
velocity is insufficient for X-ray production, and instead shows a
relatively low-ionization optical spectrum.  This is analogous to the
shock excitation in Herbig-Haro jets (Hartigan et al.\ 1990; Morse et
al.\ 1994) and may explain why we see [N~{\sc ii}] emission even
though the observed speeds imply that the shocked material should be
non-radiative.  We did not detect EFE south of the star in the RC Spec
slit aperture offset to the southwest of $\eta$ Car.  This absence is
particularly interesting, since that position coincides with a
distinct gap in the soft X-ray shell around $\eta$ Car (see Figure 3).

Finally, the `EFE' in Figure 4 may help explain the peculiar
morphology of the E Condensations (see Figs.\ 1 and 3), which show
teardrop-like shapes that sweep away from the central star.  If the E
Condensations are dense knots that are being overtaken by a much
faster and more tenuous outflowing wind or blast wave, their structure
may arise from Rayleigh-Taylor instabilities.  The detailed structure
in the E Condensations resembles some Rayleigh-Taylor instabilities in
the supernova remnant Cas A, for example (Fesen et al.\ 2001), or
simulations of wind-cloud interactions (e.g., Klein et al.\ 1994).
Note that the E Condensations do coincide with a prominent feature in
the soft X-ray shell (Figure 3).

\subsection{Sudden Chemical Enrichment?}

Of the two possible scenarios to explain the observed N and O line
intensities --- chemical abundance gradients in the dense knots vs.\
modified abundances through swept-up material --- the second seems
more likely for reasons described above.  In either case, however, the
observed abundance gradient is significant, because it indicates that
$\eta$ Car's ejection of nitrogen-rich ashes of the CNO cycle is a
{\it recent} phenomenon, occurring in just the past few thousand
years.  (The oldest of the outer condensations measured by Walborn et
al.\ (1978) have proper motions indicating ages of $\sim$10$^3$
years.)  We have discussed several interesting aspects of the observed
kinematics, but the main conclusion of this paper is as follows:

{\it Our observations suggest that diffuse gas immediately outside the
nitrogen-rich condensations seen in images of $\eta$ Car --- the
material they are running into --- has not been significantly
processed through the CNO burning cycle.}

The surrounding cocoon of normal composition material that the outer
ejecta are now plowing through (see Figure 5) probably corresponds to
part of the star's original H-rich stellar envelope.  Lamers et al.\
(2001) have inferred a similar type of abundance gradient or rapid
enrichment deduced from N/O ratios in a sample of other LBV nebulae.
Since the Outer Ejecta are so young ($\la$10$^3$ years), this
conjecture has interesting implications for stellar evolution theories
for the most massive stars.  For example, stellar evolution models for
stars above 50 to 60 M$_{\odot}$ predict that internal turbulent
mixing timescales are shorter than mass-loss timescales on the main
sequence; a star with an initial mass of 120 M$_{\odot}$ has a mixing
timescale of only 1.6 Myr compared to a mass-loss timescale of 2.4
Myr, both of which are shorter than the time spent on the main
sequence (Maeder 1982).  Rotation tends to enhance the mixing (Maeder
\& Meynet 2002; Maeder 2002), and there are indications that $\eta$
Car has significant rotation (Smith et al.\ 2003a).  Thus, we should
expect the cores and envelopes of very massive stars to be well-mixed
and to evolve quasi-homogeneously.

Instead, our observations suggest that $\eta$ Car's outer layers were
blown off the star before turbulent mixing was able to transport
nitrogen-rich CNO products to the star's surface.  This might imply
that in very massive stars with initial masses above 100 M$_{\odot}$,
the mass-loss timescale on the main sequence is shorter than the
mixing timescale between the core and outer layers, shorter than the
nuclear-burning timescale for the CNO cycle (e.g., Appenzeller 1970),
or that violent sporadic mass-loss events like $\eta$ Car's Great
Eruption play a central role in a very massive star's evolution off
the main sequence.  The cocoon around $\eta$ Car may have been created
during post-main sequence evolution immediately before its current
tenure as an extreme luminous blue variable, consistent with the idea
that $\eta$ Car may be a rare example of a post-WNL type star (Walborn
1989; see also Langer et al.\ 1994; Crowther et al.\ 1995).


The apparent abundance variations we observe in $\eta$ Car's Outer
Ejecta provide grounds for interesting though inconclusive
speculation regarding its current evolutionary state: Is there a link
between $\eta$ Car's fundamental instability and the critical point in
the star's evolution when CNO products are first exposed at the
surface?  Are catastrophic mass-loss events like the Great Eruption
responsible for removing the outer H-rich envelope to expose the CNO
ashes, or do they occur as a result of it (i.e. from a change in
opacity)?  The Great Eruption ejected several solar masses of material
from the star (Mitchell \& Robinson 1978; Hackwell et al.\ 1986; Cox
et al.\ 1995; Smith et al.\ 1998, 2003b), and therefore removed a
large fraction of the star's outer radius.  For example, models by
Guzik et al.\ (1999) predict extremely tenuous outer layers for a star
like $\eta$ Car, with 95\% of the radius containing less than 1\% of
the total mass.  It is interesting to speculate that the Great
Eruption itself, or perhaps a previous similar event, may have been
the trigger that first released the CNO ashes from the star.  Walborn
(1976) has already compared $\eta$ Car's Outer Ejecta to the
nitrogen-rich ``quasi stationary flocculi'' in the supernova remnant
Cas A (Baade \& Minkowski 1954; van den Bergh 1971; Chevalier \&
Kirshner 1978; Fesen et al.\ 1987, 2001); these existed in the
presupernova circumstellar environment and were probably ejected
shortly before the progenitor star finally exploded.  The
corresponding implications of the nitrogen-rich Outer Ejecta for
$\eta$ Car's evolutionary state and its near future are provocative.

\acknowledgements \footnotesize

We thank Mike Corcoran for supplying the Chandra imaging data used in
Figure 3, and we thank an anonymous referee for constructive comments
that improved the paper.  NOAO paid for travel to Chile and
accommodations while observing at CTIO.  Additional support was
provided by NASA grant NAG 5-12279 to the University of Colorado, and
through grants HF-01166.01-A and GO-08178.01-A from the Space
Telescope Science Institute, which is operated by the Association of
Universities for Research in Astronomy, Inc., under NASA contract NAS
5-26555.

\begin{deluxetable}{lcccccc}
\tighten
\tablewidth{0pt}
\tablecaption{Observed Emission Line Intensities\tablenotemark{a}}
\tablehead{
  \colhead{Line I.D.} &\colhead{S~Cond.} &\colhead{S~Ridge} &\colhead{W~Cond.~2} 
                      &\colhead{E~Cond.~5} &\colhead{W~Edge} &\colhead{H~{\sc ii}~reg.}
}
\startdata
$[$O~{\sc ii}] 3726+3729      &11.0	      &4.2	      &97.4	      &124	      &305	      &168       \\
H$\gamma$ 4340	              &45.9	      &47.8	      &47.2	      &47.5	      &49.2	      &45.2      \\
$[$O~{\sc iii}] 4363\tablenotemark{b} &\nodata &\nodata	      &$<$1.7	      &1.0	      &2.5	      &3.5       \\
H$\beta$ 4861   	      &100	      &100	      &100	      &100	      &100	      &100       \\
$[$O~{\sc iii}] 4959	      &$<$1.0	      &$<$0.1	      &14.7	      &30.2	      &59.3	      &60.1      \\
$[$O~{\sc iii}] 5007	      &$<$3.0	      &$<$0.3	      &49.1	      &89.6	      &182	      &185       \\
$[$N~{\sc i}] 5200	      &51.2	      &8.6	      &$<$1.7 	      &1.9	      &$<$3.3	      &1.6       \\
$[$N~{\sc ii}] 5755	      &69.2	      &12.2	      &6.8	      &14.5	      &$<$4.9	      &1.4       \\
He~{\sc i} 5876 	      &25.3	      &17.9	      &10.7	      &19.1	      &21.3	      &14.8      \\
$[$O~{\sc i}] 6300	      &(6.8)	      &(1.2)	      &$<$1.7 	      &(2.4)	      &$<$4.0	      &0.7       \\
$[$S~{\sc iii}] 6312	      &(2.2)	      &(3.2)	      &$<$1.7 	      &(2.4)	      &(5.8)	      &1.7       \\
$[$O~{\sc i}] 6364	      &(6.5)	      &(1.2)	      &$<$1.7 	      &(1.9)	      &$<$4.0	      &0.7       \\
$[$N~{\sc ii}] 6548	      & 1010	      &385	      &76.9	      &311	      &29.0	      &15.5      \\
H$\alpha$ 6563  	      &373	      &474	      &377	      &460	      &427	      &408       \\
$[$N~{\sc ii}] 6583	      & 3060	      & 1160	      &238	      &938	      &89.8	      &46.5      \\
He~{\sc i} 6678 	      &10.8	      &13.3	      &3.5	      &8.1	      &8.1	      &6.0       \\
$[$S~{\sc ii}] 6717	      &62.7	      &17.7	      &4.4	      &12.0	      &12.1	      &16.0      \\
$[$S~{\sc ii}] 6731	      &116	      &28.5	      &5.7	      &20.1	      &11.4	      &11.9      \\
He~{\sc i} 7065 	      &16.0	      &16.6	      &9.8	      &11.3	      &6.8	      &6.4       \\
$[$Ar~{\sc iii}] 7136         &7.6            &4.1            &5.7            &16.7           &22.8           &18.7      \\
$[$Fe~{\sc ii}] 7155	      &48.2           &8.5            &$<$1.7         &8.6            &$<$4.0         &4.0       \\
$[$Ca~{\sc ii}] 7291	      &49.7           &9.6            &$<$1.7         &11.4           &$<$4.0         &$<$0.1    \\
$[$Ca~{\sc ii}] 7325	      &34.7           &5.1            &$<$1.7         &12.6           &\nodata        &\nodata   \\
$[$O~{\sc ii}] 7325	      &\nodata        &\nodata        &$<$1.7         &\nodata        &11.2           &3.6       \\
$[$Ni~{\sc ii}] 7379	      &41.4           &12.6           &4.4            &14.0           &$<$4.0         &1.0       \\
$[$Ni~{\sc ii}] 7412	      &9.1            &2.0            &$<$1.7         &2.2            &$<$4.0         &0.4       \\
$[$Fe~{\sc ii}] 7452	      &15.7           &2.8            &$<$1.7         &3.3            &$<$4.0         &$<$0.1    \\
$[$Fe~{\sc ii}] 8617	      &65.9           &13.5           &3.5            &14.2           &$<$4.0         &0.3       \\
$[$S~{\sc iii}] 9069	      &42.0	      &49.4	      &21.4	      &48.5	      &(19)	      &60.4      \\
$[$S~{\sc iii}] 9532	      &103	      &146	      &70.8	      &237	      &256	      &214       \\
\enddata
\tablenotetext{a}{Quantities in parentheses are uncertain due to
blending or low signal-to-noise.}
\tablenotetext{b}{For the S Condensation and S Ridge we cannot give
useful upper limits for the [O~{\sc iii}] $\lambda$4363 intensity,
because [Fe~{\sc ii}] $\lambda$4358 dominates the emission at this
wavelength.}
\end{deluxetable}

\begin{deluxetable}{lcccccc}
\tighten
\tablewidth{0pt}
\tablecaption{Dereddened Emission Line Intensities and Derived Properties\tablenotemark{a}}
\tablehead{
  \colhead{Line I.D.} &\colhead{S~Cond.} &\colhead{S~Ridge} &\colhead{W~Cond.~2} 
                      &\colhead{E~Cond.~5} &\colhead{W~Edge} &\colhead{H~{\sc ii}~reg.}
}
\startdata
$[$O~{\sc ii}] 3726+3729      &13.3	      &5.0	      &117	      &150	      &368	      &203       \\
H$\gamma$ 4340  	      &50.2	      &52.2	      &51.6	      &52.0	      &53.9	      &49.4      \\
$[$O~{\sc iii}] 4363	      &\nodata	      &\nodata	      &$<$1.9	      &1.1	      &2.8	      &3.9       \\
H$\beta$ 4861   	      &100	      &100	      &100	      &100	      &100	      &100       \\
$[$O~{\sc iii}] 4959	      &$<$0.9	      &$<$0.08	      &14.4	      &29.6	      &58.3	      &59.1      \\
$[$O~{\sc iii}] 5007	      &$<$2.9	      &$<$0.26	      &47.8	      &87.3	      &177	      &180       \\
$[$N~{\sc i}] 5200	      &48.3	      &8.1	      &$<$1.6	      &1.8	      &$<$3.1	      &1.5       \\
$[$N~{\sc ii}] 5755	      &60.4	      &10.7	      &5.9	      &12.6	      &$<$4.3	      &1.2       \\
He~{\sc i} 5876 	      &21.8	      &15.4	      &9.2	      &16.5	      &18.3	      &12.8      \\
$[$O~{\sc i}] 6300	      &(5.6)	      &(1.0)	      &$<$1.4	      &(2.0)	      &$<$3.3	      &0.6       \\
$[$S~{\sc iii}] 6312	      &(1.8)	      &(2.7)	      &$<$1.4         &(2.0)	      &(4.8)	      &1.4       \\
$[$O~{\sc i}] 6364	      &(5.3)	      &(1.0)	      &$<$1.4	      &1.6	      &$<$3.3	      &0.6       \\
$[$N~{\sc ii}] 6548	      &816	      &312	      &62.3	      &252	      &23.5	      &12.6      \\
H$\alpha$ 6563  	      &302	      &384	      &306	      &373	      &345	      &331       \\
$[$N~{\sc ii}] 6583	      &2470	      &942	      &193	      &758	      &72.6	      &37.6      \\
He~{\sc i} 6678 	      &8.7	      &10.7	      &2.8	      &6.5	      &6.5	      &4.8       \\
$[$S~{\sc ii}] 6717	      &50.2	      &14.1	      &3.5	      &9.6	      &9.7	      &12.8      \\
$[$S~{\sc ii}] 6731	      &92.6	      &22.8	      &4.5	      &16.1	      &9.1	      &9.5       \\
He~{\sc i} 7065 	      &12.5	      &12.9	      &7.7	      &8.8	      &5.3	      &5.0       \\
$[$Ar~{\sc iii}] 7136         &5.9            &3.2            &4.4            &12.9           &17.7           &14. 5     \\
$[$Fe~{\sc ii}] 7155	      &37.3           &6.6            &$<$1.1         &6.6            &$<$2.6         &3.1       \\
$[$Ca~{\sc ii}] 7291	      &38.1           &7.4            &$<$1.1         &8.8            &$<$2.6         &$<$0.1    \\
$[$Ca~{\sc ii}] 7325	      &26.5           &3.9            &$<$1.1         &9.6            &\nodata        &\nodata   \\
$[$O~{\sc ii}] 7325	      &\nodata        &\nodata        &$<$1.1         &\nodata        &8.5            &2.7       \\
$[$Ni~{\sc ii}] 7379	      &31.5           &9.6            &3.3            &10.6           &$<$2.5         &0.8       \\
$[$Ni~{\sc ii}] 7412	      &6.9            &1.5            &$<$1.0         &1.7            &$<$2.5         &0.3       \\
$[$Fe~{\sc ii}] 7452	      &11.9           &2.1            &$<$1.0         &2.5            &$<$2.5         &$<$0.1    \\
$[$Fe~{\sc ii}] 8617	      &45.2           &9.2            &2.4            &9.7            &$<$2.1         &0.2       \\
$[$S~{\sc iii}] 9069	      &28.0	      &33.0	      &14.2	      &32.3	      &(13)	      &40.3      \\
$[$S~{\sc iii}] 9532	      &68.0	      &95.6	      &46.2	      &155	      &167	      &140       \\
\tableline
n$_e$ [S~{\sc ii}] (cm$^{-3}$) &7,100$\pm$1800 &3,600$\pm$700 &1,540$\pm$460  &4,230$\pm$1120 &470$\pm$170    &$\la$150         \\
T$_e$ [N~{\sc ii}] (K)	      &11,550$\pm$600  &8,800$\pm$300 &14,600$\pm$600 &10,150$\pm$400 &$<$25,000      &15,800$\pm$500   \\
T$_e$ [O~{\sc iii}] (K)       &\nodata	      &\nodata	      &$<$22,000      &12,400$\pm$800 &13,700$\pm$1000 &15,800$\pm$1300 \\
n(N$^0$)/n(H)  &5.9$\times$10$^{-5}$ &1.8$\times$10$^{-5}$ &$<$4.2$\times$10$^{-7}$ &2.4$\times$10$^{-6}$ &$<$7.3$\times$10$^{-7}$ &1.8$\times$10$^{-7}$ \\
n(N$^+$)/n(H)  &3.7$\times$10$^{-4}$ &2.8$\times$10$^{-4}$ &1.6$\times$10$^{-5}$    &1.5$\times$10$^{-4}$ &6.8$\times$10$^{-6}$    &2.7$\times$10$^{-6}$ \\
n(O$^0$)/n(H)  &1.5$\times$10$^{-5}$ &6.6$\times$10$^{-6}$ &$<$1.6$\times$10$^{-6}$ &3.2$\times$10$^{-6}$ &$<$4.5$\times$10$^{-6}$ &5.0$\times$10$^{-7}$ \\
n(O$^+$)/n(H)  &5.7$\times$10$^{-6}$ &5.4$\times$10$^{-6}$ &1.4$\times$10$^{-5}$    &4.0$\times$10$^{-5}$ &4.5$\times$10$^{-5}$    &1.5$\times$10$^{-5}$ \\
n(O$^{+2}$)/n(H) &$<$6.4$\times$10$^{-7}$ &$<$1.4$\times$10$^{-7}$ &5.4$\times$10$^{-6}$ &1.6$\times$10$^{-5}$ &2.4$\times$10$^{-5}$ &1.7$\times$10$^{-5}$ \\
n(N+O)/n(H)\tablenotemark{b}    &4.5$\times$10$^{-4}$ &3.1$\times$10$^{-4}$ &3.7$\times$10$^{-5}$ &2.1$\times$10$^{-4}$ &8.1$\times$10$^{-5}$ &3.5$\times$10$^{-5}$ \\
n(N)/n(O)\tablenotemark{c}      &22$\pm$7             &24$\pm$6             &0.8$\pm$0.3          &2.6$\pm$0.8          &0.1$\pm$0.05         &0.1$\pm$0.03         \\
\enddata
\tablenotetext{a}{\scriptsize Quantities in parentheses are uncertain due to
blending or low signal-to-noise.  Intensities have been corrected with
the reddening law of Cardelli et al.\ (1989) using $E(B-V)$=0.2.}
\tablenotetext{b}{\scriptsize This is the observed quantity
n(N$^0$+N$^+$+O$^0$+O$^+$+O$^{+2}$)/n(H), and is not the total N+O
abundance relative to H.}  
\tablenotetext{c}{\scriptsize This is the observed ratio
n(N$^0$+N$^+$)/n(O$^0$+O$^+$+O$^{+2}$), and is not the total N/O
ratio.}
\end{deluxetable}

\begin{figure}
\epsscale{0.9}
\plotone{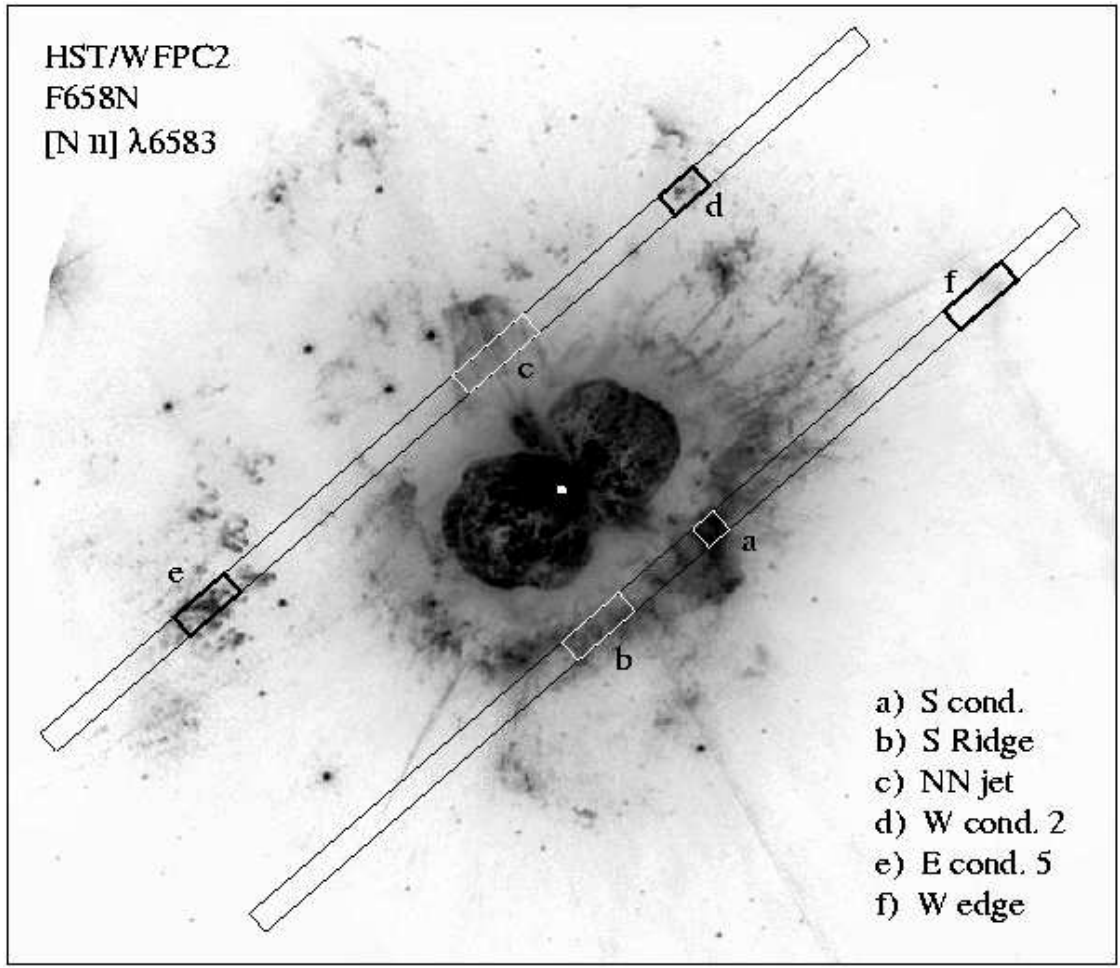}
\caption{{\it HST}/WFPC2 image of $\eta$ Carinae, the Homunculus, and
the Outer Ejecta seen in the F658N filter transmitting mostly [N~{\sc
ii}] $\lambda$6583 emission (see Morse 1999).  Long-slit aperture
positions for the RC Spec observations are indicated, as are
sub-apertures for various features labeled $a$-$f$ whose extracted 1-D
spectra are shown in Figure 2$a$-$f$, respectively.  The fastest
features are Doppler-shifted out of the filter.  This image has been
lightly processed with unsharp-masking to enhance contrast in the huge
dynamic range.}
\end{figure}

\begin{figure}
\epsscale{0.9}
\plotone{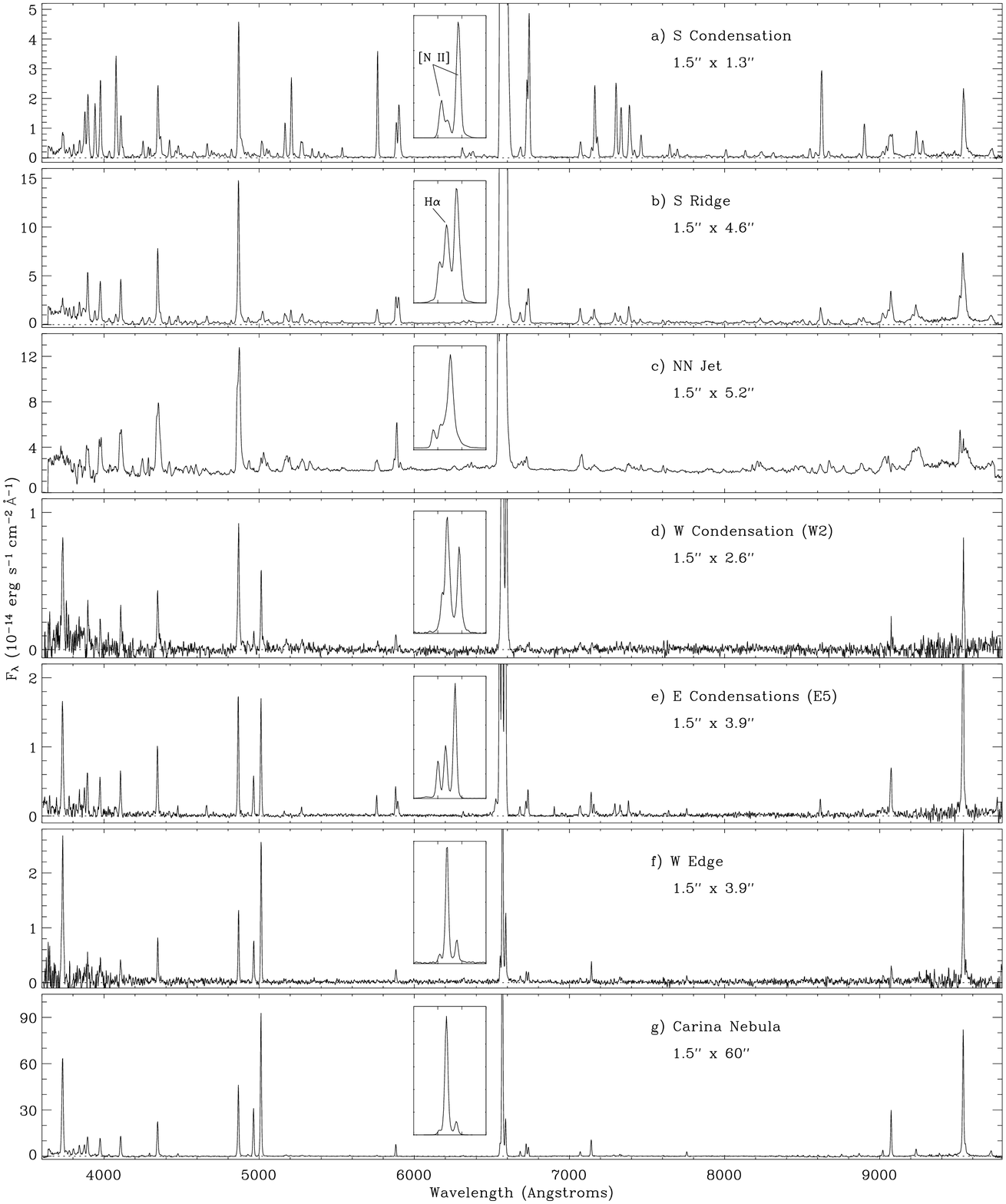}
\caption{ Ground-based spectra of various features in the Outer Ejecta
of $\eta$ Carinae.  Panels $a$ - $f$ correspond to features labeled in
Figure 1.  Panel $g$ is the average spectrum of the background Carina
Nebula H~{\sc ii} region near $\eta$ Car.}
\end{figure}

\begin{figure}
\epsscale{0.9}
\plotone{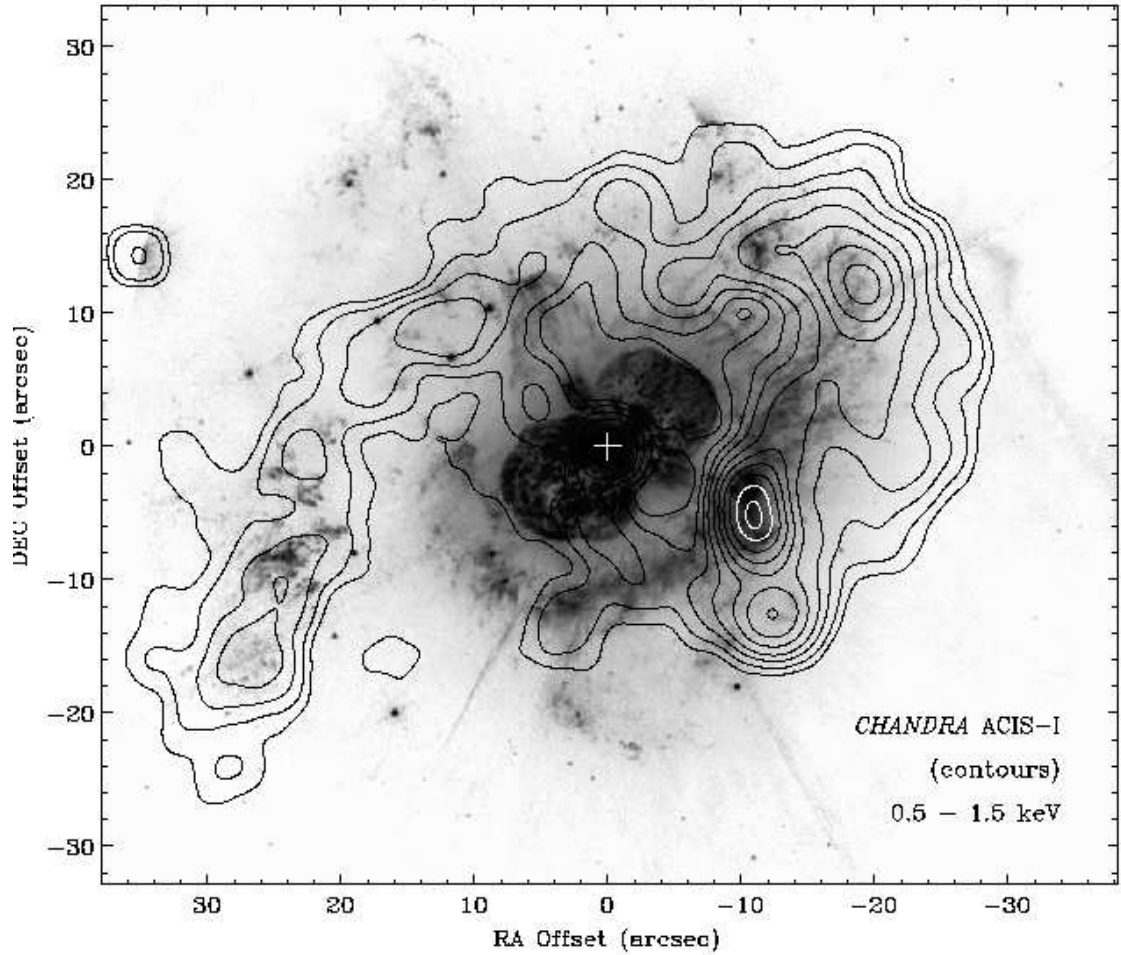}
\caption{{\it HST}/WFPC2 image from Figure 1 with contours of soft
X-ray emission observed by {\it CHANDRA} ACIS-I superposed.  The X-ray
flux is integrated from roughly 0.5 to 1.5 keV, and the X-ray image
has been adaptively smoothed.  The two images were spatially aligned
by matching the position of the central star at visual wavelengths
with the position of the bright hard X-ray source.  The {\it Chandra}
imaging data were kindly provided to us by M.\ Corcoran (see Seward et
al.\ 2001).}
\end{figure}

\begin{figure}
\epsscale{0.5}
\plotone{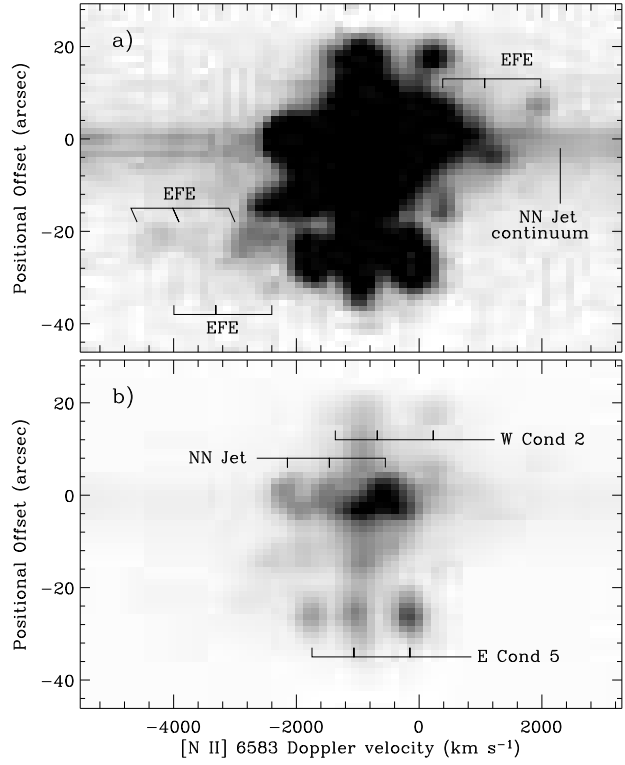}
\caption{A portion of the 2-D long-slit spectrum for the slit position
centered $\sim$10$\arcsec$ northeast of $\eta$ Car (see Figure 1).
Doppler velocities on the horizontal axis are indicated for the
[N~{\sc ii}] $\lambda$6583 line, but several labels identify all three
lines ([N~{\sc ii}] $\lambda$6548, H$\alpha$, and [N~{\sc ii}]
$\lambda$6583) for various features.  Panel (a) has the intensity
scale set to show the faintest emission features. Panel (b) shows the
same data with nearly the full intensity range.  The `extremely fast
ejecta' discussed in \S 4.3 are labeled as `EFE'.  Note that the E
Cond.\ 5 has a radial velocity of about $-$140 km s$^{-1}$, consistent
with independent estimates from echelle spectra (Meaburn et al.\
1996).}
\end{figure}

\begin{figure}
\epsscale{0.6}
\plotone{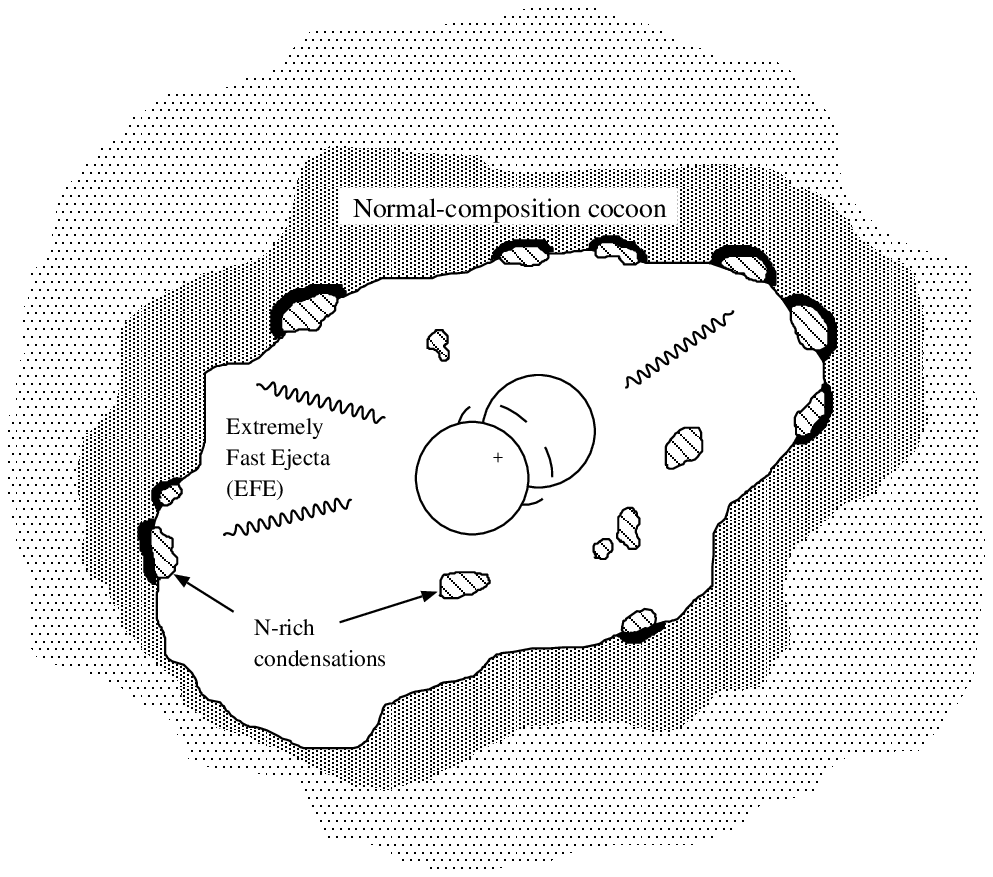}
\caption{Schematic cartoon depicting the Outer Ejecta of $\eta$
Carinae.  The star has ejected several nitrogen-rich condensations
that are expanding away from the bipolar Homunculus.  Some of these
N-rich condensations are expanding in a cavity, and some are colliding
with a normal-composition cocoon deposited by previous stellar-wind
mass loss.  This cocoon is, in turn, expanding into the surrounding
H~{\sc ii} region.  The cavity inside this normal composition cocoon
and outside the Homunculus is partly filled with the extremely fast
ejecta (EFE; see Figure 4).  Shocks at the inside edge of the cocoon
give rise to soft X-ray emission (see Figure 3).}
\end{figure}


\begin{references}

\reference{} Appenzeller, I.\ 1970, A\&A, 9, 216

\reference{} Baade, W., \& Minkowski, R.\ 1954, ApJ, 119, 206

\reference{} Brooks, K.J., Burton, M.G., Rathborne, J.M., Ashley,
M.C.B., \& Storey, J.W.V.\ 2000, MNRAS, 319, 95

\reference{} Cardelli, J.A., Clayton, G.C., \& Mathis, J.S.\ 1989,
ApJ, 345, 245

\reference{} Chevalier, R.A., \& Kirshner, R.P.\ 1978, ApJ, 219, 931

\reference{} Chevalier, R.A., \& Raymond, J.C.\ 1978, ApJ, 225, L27

\reference{} Chlebowski, T., Seward, F.D., Swank, J., \& Szymkowiak,
A.\ 1984, ApJ, 281, 665

\reference{} Cox, P., et al.\ 1995, A\&A, 297, 168

\reference{} Crowther, P.A., Smith, L.J., Hillier, D.J., \& Schmutz,
W.\ 1995, A\&A, 293, 427

\reference{} Davidson, K., Dufour, R.J., Walborn, N.R., \& Gull, T.R.\
1986, ApJ, 305, 867

\reference{} Davidson, K., Walborn, N.R., \& Gull, T.R.\ 1982, ApJ,
254, L47

\reference{} Dufour, R.J.\ 1989, RevMexAA, 18, 87

\reference{} Dufour, R.J., Glover, T.W., Hester, J.J., Currie, D.G.,
van Orsow, D., \& Walter, D.K.\ 1997, in ASP Conf.\ Ser.\ 120,
Luminous Blue Variables: Massive Stars in Transition, ed. A.\ Nota \&
H.J.G.L.M.\ Lamers (San Francisco: ASP), 255

\reference{} Fesen, R.A., Becker, R.H., \& Blair, W.P.\ 1987, ApJ, 313, 378

\reference{} Fesen, R.A., Becker, R.H., \& Goodrich, R.W.\ 1988, ApJ, 329, L89

\reference{} Fesen, R.A., Morse, J.A., Chevalier, R.A., Borkowski,
K.J., Gerardy, C.L., Lawrence, S.S., \& van den Bergh, S.\ 2001, AJ,
122, 2644

\reference{} Guzik, J.A., Cox, A.N., \& Despain, K.M.\ 1999, in ASP
Conf.\ Ser.\ 179, Eta Carinae at the Millenium, ed.\ J.A.\ Morse,
R.M.\ Humphreys, \& A.\ Damineli (San Francisco: ASP), 347

\reference{} Hackwell, J.A., Gehrz, R.D., \& Grasdalen, G.L.\ 1986, ApJ, 311, 380

\reference{} Hartigan, P., Raymond, J., \& Meaburn, J.\ 1990, ApJ,
362, 624

\reference{} Hummer, D.G., \& Storey, P.J.\ 1987, MNRAS, 224, 801

\reference{} Klein, R.I., McKee, C.F., \& Colella, P.\ 1994, ApJ, 420, 213

\reference{} Lamers, H.J.G.L.M., Nota, A., Panagia, N., Smith, L.J.,
\& Langer, N.\ 2001, ApJ, 551, 764

\reference{} Langer, N., Hamann, W.R., Lennon, M., Najarro, F.,
Pauldrach, A.W.A., \& Puls, J.\ 1994, A\&A, 290, 819

\reference{} Maeder, A.\ 1982, A\&A, 105, 149

\reference{} Maeder, A.\ 2002, A\&A, 392, 575

\reference{} Maeder, A., \& Meynet, G.\ 2002, ARA\&A, 38, 143

\reference{} Meaburn, J., Wolstencroft, R.D., \& Walsh, J.R.\ 1987,
A\&A, 181, 333

\reference{} Meaburn, J., Gehring, G., Walsh, J.R., Palmer, J.W.,
Lopez, J.A., Bryce, M., \& Raga, A.C.\ 1993, A\&A, 276, L21

\reference{} Meaburn, J., Boumis, P., Walsh, J.R., Steffen, W.,
Holloway, A.J., Williams, R.J.R., \& Bryce, M.\ 1996, MNRAS, 282, 1313

\reference{} Meaburn, J., Bryce, M., \& Holloway, A.J.\ 1995, A\&A,
299, L1

\reference{} Mitchell, R.M., \& Robinson, G.\ 1978, ApJ, 220, 841

\reference{} Morse, J.A.\ 1999, in ASP Conf.\ Ser.\ 179, Eta Carinae
at the Millenium, ed.\ J.A.\ Morse, R.M.\ Humphreys, \& A.\ Damineli
(San Francisco: ASP), 13

\reference{} Morse, J.A., Davidson, K., Bally, J., Ebbets, D., Balick,
B., \& Frank, A.\ 1998, AJ, 116, 2443

\reference{} Morse, J.A., Hartigan, P., Heathcote, S., Raymond, J., \&
Cecil, G.\ 1994, ApJ, 425, 738

\reference{} Morse, J.A., Kellogg, J.R., Bally, J., Davidson, K.,
Balick, B., \& Ebbets, D.\ 2001, ApJ, 548, L207

\reference{} Pittard, J.M., \& Corcoran, M.F.\ 2002, A\&A, 383, 636

\reference{} Seward, F.D., Butt, Y.M., Karovska, M., Prestwich, A.,
Schlegel, E.M., \& Corcoran, M.F.\ 2001, ApJ, 553, 832

\reference{} Shaw, R.A., \& Dufour, R.J.\ 1994, in ASP Conf.\ Ser.\
61, Astronomical Data Analysis Software and Systems {\sc iii}, ed.\
D.R. Crabtree, R.J.\ Hanisch, \& J.\ Barbes (San Francisco: ASP), 327

\reference{} Smith, N.\ 2002, MNRAS, 331, 7

\reference{} Smith, N., Davidson, K., Gull, T.R., Ishibashi, K., \&
Hillier, D.J.\ 2003a, ApJ, 586, 432

\reference{} Smith, N., Gehrz, R.D., \& Krautter, J.\ 1998, AJ, 116, 1332

\reference{} Smith, N., Gehrz, R.D., Hinz, P.M., Hoffmann, W.F., Hora,
J.L., Mamajek, E.E., \& Meyer, M.R.\ 2003b, AJ, 125, 1458

\reference{} Smith, R.G.\ 1987, MNRAS, 277, 943

\reference{} Sonneborn, G., Fransson, C., Lundqvist, P., Cassatella,
A., Gilmozzi, R., Kirshner, R.P., Panagia, N., \& Wamsteker, W.\ 1997,
ApJ, 477, 848

\reference{} Thakeray, A.D.\ 1950, MNRAS, 110, 524

\reference{} van den Bergh, S.\ 1971, ApJ, 156, 457

\reference{} Walborn, N.R.\ 1976, ApJ, 204, L17

\reference{} Walborn, N.R.\ 1989, in Physics of Luminous Blue
Variables, ed.\ K.\ Davidson, A.F.J.\ Moffat, \& H.J.G.L.M.\ Lamers
(Dordrecht:\ Kluwer), 251

\reference{} Walborn, N.R.\ 1995, Rev.\ Mexicana A\&A, Ser.\ Conf., 2, 51

\reference{} Walborn, N.R., \& Blanco, B.M.\ 1988, PASP, 100, 797

\reference{} Walborn, N.R., Blanco, B.M., \& Thackeray, A.D.\ 1978,
ApJ, 219, 498

\reference{} Weis, K., Duschl, W.J., \& Bomans, D.J.\ 2001, A\&A, 367,
566


\end{references}
\end{document}